\documentstyle[preprint,eqsecnum,epsfig,prd,aps,floats]{revtex} 

\begin{document}

\newcommand{\nl}{\nonumber \\}
\newcommand{\eq}[1]{Eq.~(\ref{#1})}

\preprint{\vbox{ \tighten {
		\hbox{PUPT-1951}
		}  }}  
  
\title{Compactified NCOS and duality}
  
\author{Martin Gremm}
  
\address{Joseph Henry Laboratories, Princeton University, Princeton, NJ 08544}
 
\maketitle  
 
\begin{abstract} 
\tighten{ 
We study four-dimensional $U(1)$ on a non-commutative $T^2$ with rational
$\Theta$. This theory has dual descriptions as ordinary SYM or as NCOS. We 
identify a set of massive non-interacting KK states in the SYM theory and
track them through the various dualities. They appear as stretched strings
in the non-commutative $U(1)$ providing another example of the IR/UV mixing
in non-commutative field theories. In the NCOS these states appear as D-strings
with winding and momentum. They form an unconventional type of 1/4 BPS state
with the 3-brane. To obtain a consistent picture of S-duality for compactified
theories it is essential to keep track of both the NS and the RR B-fields.
} 
\end{abstract} 
 
\newpage

\section{Introduction} 
 
Non-commutative field theories arise as world volume theories on D-branes
in constant B-field backgrounds \cite{douglas,sw}. One may hope that a better 
understanding of these field theories will lead to a better understanding
of these backgrounds in string theory.

Apart from that non-commutative 
field theories have many unusual properties, that make them interesting 
in their own right. One of the more intriguing properties of non-commutative
Yang-Mills theories is S-duality \cite{sdual,sdual2,sdual3,sdual4}.
The S-dual of four-dimensional NCYM turns out
to be a non-commutative open string theory (NCOS). Such theories are very
unusual, since generally poles in open string amplitude require the inclusion
of closed strings to ensure unitarity. The only other known example of
consistent string theories without gravity are the Little String
Theories \cite{lst},
but unlike the NCOS, these theories have no weak coupling
limit which makes them hard to study (see however \cite{lst2}).
The NCOS on the other hand can have 
arbitrarily small coupling and we can study it in string perturbation theory. 

In \cite{igor,om} is was pointed out that in the 1+1 dimensional case the
NCOS is dual to an ordinary maximally supersymmetric gauge theory with one unit
of electric flux turned on (see also \cite{also,also2}).
This provides one example where the exotic NCOS has an equivalent
description in terms of an ordinary commutative field theory. In this paper
we provide another example of this type. More precisely, we study
four-dimensional $U(1)$ compactified on a non-commutative $T^2$ \cite{ft}.
If we choose the dimensionless 
non-commutativity parameter $\Theta = 1/s$, where $s$ is an integer, this 
theory has an equivalent description as a commutative $U(s)$ theory with
one unit of magnetic flux. On the other hand, the original non-commutative
$U(1)$ has an S-dual description as a four-dimensional NCOS on a torus. In
the 1+1 dimensional case there were only two descriptions of the theory, NCOS
and ordinary field theory. In our case there is a third as a non-commutative
field theory. 

Some features of these theories can be understood from the underlying
string description \cite{douglas,sw}.
The non-commutative $U(1)$ theory can be realized as
the decoupling limit of a single 3-brane on $T^2$ with $s$ units of NS B-flux. 
This B-flux through the torus induces $s$ D-string
charges in the non-compact directions. Note that these D-strings are a 
source for the RR B-field, so in general both the NS and the RR B-field are
turned on. T-duality on the torus turns the induced D-strings into D3 branes
and the original 3-brane into a D-string. 
Thus we find a $U(s)$ gauge theory with one unit of magnetic flux.
We review the field theory picture of this duality in the next section. 

In the commutative Yang-Mills description the gauge group factorizes as
$U(s)=(U(1)\times SU(s))/Z_s$, and the $U(1)$ is free. Since the theory is 
compactified on $T^2$, we expect a KK tower of massive non-interacting 1/4
BPS states from the reduction of the $U(1)$. It is interesting to identify
these states in the other two descriptions of this theory as well. 

The KK states of the commutative theory turn into states 
carrying both momentum and winding in the dual description as a non-commutative
$U(1)$. According to \cite{dipole} an open string state moving in one of the
non-commutative directions has a finite extent in the other. This statement
is strictly true in the $\alpha^\prime\to 0$ limit, which eliminates the
oscillator contributions. However, this is the field theory limit we want
to take in the end, so thinking of these open strings as objects with a
well-defined length makes sense. 

It turns out that the states dual to the KK states of the $U(1)$
carry precisely the required amount of momentum in one
compact direction to wrap around the other an integer number of times. 
This provides another example of the IR/UV connection in non-commutative
field theory \cite{iruv}.
Since these open strings wrap a compact direction an integer number of times,
we expect that they can turn into closed strings. In Section \ref{str} we
show that there are indeed closed strings with degenerate mass. However it
turns out that all open string states that have integer winding around the
compact directions decouple from the degrees of freedom of the non-commutative
$U(1)$. We show that this is the case by computing the tree level scattering
amplitude for four open string states. If one or more of the string states
can turn into a closed string, the amplitude vanishes in the field theory 
limit. 

In Sect.~\ref{dstr} we discuss the S-dual picture. Since the BPS states of 
interest could be identified with stretched open strings before S-duality,
they turn into D-strings in the NCOS. We show that D-strings with
the appropriate winding and momentum quantum numbers form a marginally bound
1/4 BPS state with the D3-brane, while all other D-strings form real bound
states. These 1/4 BPS states are specific to this theory and do not exist
in general. They appear to be a new type of 1/4 BPS state, since
there does not seem to be a simple way to dualize them into the familiar D0-D4
system without any fluxes turned on.

In order for the masses of these states to match the masses of the 
BPS states before S-duality, it is crucial that the NS flux through the 
torus before S-duality turns into RR flux after the duality transformation. 
In the original flat space version of this S-duality the RR fields were set to
zero \cite{sdual}, but after compactification this is no longer a valid
approach, since the winding states are sensitive to these fluxes.

\section{Non-commutative $U(1)$ and its Morita duals}

We consider four-dimensional $U(1)$ theories on a non-commutative $T^2$. 
These theories can be labeled by their gauge coupling, the size of the $T^2$,
and by the non-commutativity parameter, $\Theta = \theta/(2\pi R^2)$. Here
$T^2$ is defined by the identifications $x_{2,3} \to x_{2,3} + 2\pi R$,
$[x_2,x_3]=i\theta$, and the other two directions are uncompactified.

The tree level action for this theory is given by \cite{sw}
\begin{equation}
S = \frac{1}{g_{YM}^2} \int d^4x (\hat{F}_{ij}+\Phi_{ij})
(\hat{F}^{ij}+\Phi^{ij}),
\end{equation}
and the one-loop action was recently calculated in \cite{trekkies}.
We will be interested in theories with $\Phi=0$ and $\Theta=1/s$. The first
choice is for convenience, but theories with rational $\Theta$ are 
qualitatively different than theories with irrational $\Theta$. This 
becomes clearer if one studies suitable Morita equivalent descriptions of
this theory. Morita equivalence can be studied entirely within the field 
theory or it can be viewed as the remnant of the T-duality group of the 
underlying string realization of these theories \cite{ft}. 

For our present purposes it is sufficient to know that this theory has a
$SL(2,{\bf Z}) \times SL(2,{\bf Z})$ duality group, which is the T-duality
group of the $T^2$.  We will be interested in
the $SL(2,{\bf Z})$ that acts on the non-commutativity parameter as
\begin{equation}
\Theta \to \tilde{\Theta} = \frac{ c+d\Theta}{a+b\Theta},
\end{equation}
where $a,b,c,d$ define an $SL(2,{\bf Z})$
matrix. If $\Theta$ is rational this allows us to trade our original theory
for an equivalent theory with $\tilde{\Theta}=0$, i.e.~a commutative theory. 
The other parameters
of the theory also transform under this $SL(2,{\bf Z})$:

\begin{eqnarray}\label{trans}
\tilde{\Phi} = (a + b \Theta) \Phi - b(a+b\Theta),
&\quad& \tilde{R} = (a+b\Theta)R \\ \nonumber
\tilde{m} = a m + b N, &\quad& \tilde{N} = c m + d N
\end{eqnarray}
Here $m$ is the number of magnetic fluxes and $N$ the rank of the gauge group.
In our case the parameters of the original theory are $\Phi = m = 0$, $N=1$, 
$\Theta = 1/s$, and the $SL(2,{\bf Z})$ matrix that transforms to a commutative 
theory is given by $a=0$, $b=-c=1$, $d = s$. The dual theory has
$\tilde{\Phi} = -1/s$, $\tilde{m} = 1$, $\tilde{N}=s$, $\tilde{\Theta}=0$.
These
parameters define a commutative $U(s)$ gauge theory with one unit of magnetic
flux, compactified on a torus with radii $\tilde{R} = R/s$. 

This theory has been studied in great detail \cite{ft}. Here we just note that
the $U(s)$ gauge group factorizes into $(U(1)\times SU(s))/Z_s$ and that the
$U(1)$ is free at all energies. This in turn implies that there should be
a tower of non-interacting KK states from reducing the $U(1)$ on the torus. 
We expect these states to be BPS states with masses given by 
\begin{equation}\label{u1mass}
M^2 = \frac{s^2}{R^2} (n_2^2+n_3^2),
\end{equation}
where $n_{2,3}$ are the momentum quantum numbers in the two compact directions. 

The mass formula for the BPS states of NCYM compactified on
$T^2\times S^1$ is well known \cite{ft},
\begin{equation}\label{bps}
M = \frac{g_{YM}^2}{2V {\cal N}} (w^i+\Theta^{ij}n_j)^2 +
\frac{V}{4g_{YM}^2 {\cal N}} (m_{ij}+{\cal N} \Phi{ij})^2 +
\frac{1}{\cal N} \sqrt{ \frac{k_i^2}{\Sigma_i^2}},
\end{equation}
where $k_i = {\cal N} n_i - m_{ij} (w^j+\Theta^{jk}n_k)$,
${\cal N} = N + \frac{1}{2} m_{ij}\Theta^{ij}$, $V$ is the volume of
$T^2\times S^1$ and $\Sigma_i$ are the radii of the corresponding circles.
In these expressions $w^i$ are electric flux
quanta and $n_i$ are momentum quantum numbers in the compact directions.
The magnetic flux mentioned above is given by $m_{23}=m$.

We can readily identify the KK tower of $U(1)$ states in this expression by
setting the parameters in \eq{bps} to the values corresponding to the 
commutative $U(s)$ theory. With $\tilde{\Theta}=0$, $\tilde{w}_i=0$, 
$\tilde{N}=s$, $\tilde{\Phi}=-1/s$, $m=m_{23}=1$, $n_1=0$ and 
$\Sigma_{2,3}=\tilde{R}= R/s$, the BPS mass formula reduces to the \eq{u1mass}. 

It is interesting to trace these states to the non-commutative
description as well. Using the transformations \eq{trans}, we can transform
the KK states of the commutative theory 
into the corresponding states of the non-commutative
gauge theory. In the non-commutative case we have $\Theta=1/s$, $\Phi=0$,
$m=0$, but $w^i = \tilde{n}^i$ and $n_i =-s \epsilon_{ij} \tilde{n}^j$,
so these states carry multiples on $s$ units of momentum in the compact
directions. Inserting these quantum numbers into the BPS mass formula we
verify that these states have
the same mass as the states in the commutative description, which is of course
a reflection of the fact that these theories are Morita equivalent. While it
was straightforward to see that these states do not interact in the commutative
description, it is slightly less obvious in the non-commutative case. In the
next section we show that these states do in fact decouple by studying the 
embedding of this theory in string theory.  This may seem like overkill since
one can show decoupling of these states within the field theory by expanding
the gauge fields into momentum eigenfunctions on the torus. The commutator
term in the action vanishes for gauge fields carrying $s$ units of momentum,
which implies that these fields are free. However, the string description will
be useful when we discuss the S-dual description of this theory. 

\section{The string description of non-commutative $U(1)$}\label{str}

The theories discussed in the previous section can be obtained from string
theory by taking the decoupling limit \cite{sw} of the world volume theory
on a 3-brane in a constant NS B-field. We will adopt the notation 
and conventions of \cite{sdual}, since we want to discuss the S-dual of 
these theories in the next section. Before we identify the decoupled 
$U(1)$ states in the string description of these theories, we give a very 
brief review of the decoupling limit \cite{sw} as it applies here. 

To define a non-commutative gauge theory as a limit of world volume theory
of a 3-brane in string 
theory, we send $\alpha^\prime \to 0$ while keeping the NS B-field, the
open string metric, $G^{MN}$, and the open string coupling constant, $G_o^2$, 
fixed. The open string metric is taken to be the flat 
Minkowski metric and the NS B-field is non-zero, $B_{23}=-B_{32}=1/\theta$.
In order to keep the open string quantities finite, we have to 
scale the closed string metric
$g_{ij} = (2\pi\alpha^\prime/\theta)^2\delta_{ij}$, $i,j=2,3$, and the 
closed string coupling as $g_s= G_o^2 (2\pi\alpha^\prime/\theta)$. The
decoupled theory of the massless open strings on a 3-brane is then 
a $U(1)$  gauge theory with non-commutativity in the 2-3 plane. 
To obtain the theory discussed in the previous section we take the decoupling
limit of a single 3-brane, identify $x_{2,3}\to x_{2,3} + 2\pi R$, and
put $\theta/(2\pi R^2) = \Theta = 1/s$.

For later reference we note that turning on the NS B-field in the presence
of a 3-brane induces D-string charge on it. This in turn is a source for the
RR B-field, so a non-zero $B^{NS}_{23}$ induces a non-zero $B^{RR}_{01}$ as
can be seen e.g.~from the supergravity solutions \cite{sugra} or from the
Chern-Simons term in the 3-brane action.
The RR field will not affect the analysis in this section, but it will play 
a role when we discuss the S-dual theory.

It turns out to be easy to identify the massive non-interacting states
discussed in the previous section. Consider an open string state with
momentum $p^j=n^j/R$, $j = 2,3$. According to \cite{dipole}, a string moving in
one non-commutative direction has a finite length in the orthogonal direction,
$\Delta x_i = \theta_{ij} p^j$. Using $\theta_{ij} = (2\pi R^2/s)\epsilon_{ij}$,
we see that an open string with $s$ units of momentum in one compact
direction is long enough to wind around the other compact direction once. 
Note that this works only if $s$ in an integer. 

It is tempting to identify these states with the non-interacting massive
states discussed in the previous section, but there are several problems
with that. One can object that there does not seem to be anything special
about these open string states, so one should expect them to behave in
the same way as other open string states with momenta that are not multiples
of $s$. These states do not in general decouple in the $\alpha^\prime\to 0$
limit, since they give rise to the interacting $U(1)$ degrees of freedom. 

Another problem is that the endpoints of open strings with $s$ units of momentum
coincide. It seems likely that they can fuse, turning the open string into a
closed string. However, closed strings are expected to decouple in the limit of
\cite{sw}. We will discuss the closed string spectrum first and then argue that
open strings with momenta that are multiples of $s$ decouple.

The mass
of a closed string state is given by
\begin{equation}
m^2 = \frac{1}{\alpha^{\prime 2}} g_{ij} (v^iv^j+w^iw^jR^2) +
\frac{2}{\alpha^\prime}(N+\tilde{N}-2)
\end{equation}
where $g_{ij}$ is the closed string
metric in the compact directions, $w^i$ is the winding number and
\begin{equation}
v^i = \alpha^\prime g^{ij}\left( \frac{n_j}{R} + 2\pi B_{jk} w^k R\right)
\end{equation}
In the $\alpha^\prime\to 0$ limit, closed string states with $v^i \neq 0$
or states with excited oscillators become infinitely massive. Only states
with $v^i =0$ and no string excitations survive the decoupling limit.
Using $B_{jk} = (1/\theta)\epsilon_{jk} = s/(2\pi R^2) \epsilon_{jk}$ we
find that states satisfying $n_j + s \epsilon_{jk} w^k = 0$ remain at
finite mass in the decoupling limit.
Since both $n_j$ and $w^k$ are 
integers, this equation has solutions only if $s$ is an integer (or rational
number). It is easy to check that these states have the same mass as the
open string states discussed above and the same 
quantum numbers as the non-interacting states we obtained using Morita
equivalence in the previous section.

In order to show that the open string states with momenta that are multiples
of $s$ decouple we consider the scattering amplitude for four open string
states on the 3-brane. In the commutative case the answer is well known
\cite{scatter} and and can be obtained from 
the type I four-point amplitude of open
string states (see e.g.~\cite{book}). The main difference between the
commutative and non-commutative cases are some additional phases that
appear when the positions of vertex operators are switched
\cite{phases,dipole,igor}. 

Before taking the $\alpha^\prime\to 0$ limit the boundary correlation function
in the presence of an NS B-field reads 
\begin{equation}
\langle x^i(\tau)x^j(\tau^\prime) \rangle = -2\alpha^\prime G^{ij} +
\frac{i}{2} \theta^{ij} \epsilon(\tau-\tau^\prime).
\end{equation}
The $l$-th  vertex operator contains a factor $\exp(i k_l \cdot x(\tau))$,
where $k_l$, $l=1,\ldots,4$ denotes the momentum of the $l$-th open string
state.
Moving the $l$-th vertex operator past the $m$-th generates a 
phase $\exp( i k_l \theta k_m)$.
The complete scattering amplitude for four open string states is a sum 
of six terms that differ by the ordering or the vertex operators on the 
boundary of the worldsheet. We write schematically

\begin{eqnarray}\label{amp}
A &=& \{1,2,3,4\} + e^{i \phi_2} \{4,2,3,1\} + e^{i \phi_3} \{2,4,3,1\} \\
&&+ e^{i \phi_4}\{1,3,2,4\} + e^{i \phi_5} \{4,3,2,1\} +
e^{i \phi_6} \{1,3,4,2\}, \nonumber
\end{eqnarray}
where the numbers indicate the ordering of the vertex operators on the
boundary of the world sheet and the phases are given by 
\begin{eqnarray}
\phi_2 =  k_1 \theta k_4 \quad
\phi_3 =  k_1 \theta k_4  +  k_2\theta k_4 \\
\phi_4 =  k_2 \theta k_3 \quad
\phi_5 =  k_2 \theta k_3 +  k_1 \theta k_4\quad
\phi_6 =  k_2 \theta k_3 +  k_4 \theta k_3.\nonumber
\end{eqnarray}

Recall that the $2,3$ components of the external momenta are quantized in units
of $1/R$ and that $\theta_{ij} = (2\pi R^2/s)\epsilon_{ij}$. 
To show that states with momenta that are multiples of $s$ decouple, we need
to show that the all amplitudes involving one or more of these states vanish.
We take  the $2,3$ components of $k_4$ to be given by $q_2 s/R$
and $q_3 s/R$ respectively, and leave the other momenta unspecified. This
guarantees that the phases $\phi_2$ and $\phi_3$ are 
integer multiples of $2\pi$, i.e.~trivial phases. The remaining phases
are all given by $i k_2 \theta k_3$ up to integer multiples of $2\pi$. The
first three terms in \eq{amp} give the same contribution as the last three
up to the over all phase, so we find that for these states the amplitude is
related to the commutative amplitude by 
\begin{equation}
A_{NC} = \frac{1}{2}\left( 1+ e^{i\phi_4}\right) A_C.
\end{equation}
However, the commutative amplitude vanishes in the $\alpha^\prime\to 0$ limit,
reflecting the fact that there are no interactions in a commutative $U(1)$.
This implies that any amplitude with a state that carries momenta that are
multiples of $s/R$ vanishes. This is not true in general. If the phases 
multiplying the various terms in \eq{amp} are all different, the 
$\alpha^\prime\to 0$ limit gives a non-vanishing result.  Note again, that 
this cancellation of the phases is possible only if $s$ is an integer. 

\section{The S-dual picture}\label{dstr}

Having identified the non-interacting massive states in the non-commutative 
$U(1)$ theory as strings with specific winding and momentum quantum numbers,
we can now proceed to search for these states in the S-dual NCOS description
\cite{sdual,sdual2}. 

Before delving into details, let us outline the qualitative picture of both
the non-commutative $U(1)$ and its S-dual. Turning on the NS B-field induces
D-string charge on the 3-brane. If $\Theta=1/s$ the induced charge corresponds
to $s$ D-strings, and these branes are a source for the RR B-field. 
Following \cite{sdual}, we can S-dualize the non-commutative $U(1)$ theory.
This turns the induced D-strings into fundamental strings corresponding to
electric flux on the 3-brane, and exchanges the role of the NS and RR B-fields.
In the non-commutative $U(1)$ both $B^{NS}_{23}$ and $B^{RR}_{01}$
were non-zero.
In the S-dual theory we have $B^{RR}_{23}=-1/\theta$ and $B_{01}^{NS}$,
or, equivalently, the electric flux on the 3-brane can be computed as follows.
The electric flux is given by the solution of
\begin{equation}
\frac{\sqrt{-\hat{g}}}{\hat{g}_{str}} \frac{ \hat{g}^{00} \hat{g}^{11} F_{01} }
 {\sqrt{ 1+(2\pi\alpha^\prime)^2 \hat{g}^{00}\hat{g}^{11} F_{01}^2 }} = B_{23} =
\frac{s}{2\pi R^2},
\end{equation}
where the closed string metric is given by
$\hat{g}_{11}=-\hat{g}_{00} = \theta/(2\pi\alpha^\prime G_o^2)$ and
$\hat{g}_{22}=\hat{g}_{33}= 2\pi\alpha^\prime/(G_o^2\theta)$.
The hat over symbols denotes the S-dual quantities\footnote{The symbols 
without hats in the first part of the paper correspond to the
primed quantities in \cite{sdual}, and the symbols with hats are the unprimed
variables in \cite{sdual}.}.
This gives using $\hat{g}_{str} = R^2/(\alpha^\prime G_o^2 s)$
\begin{equation}
F_{01} =
\frac{F_{01}^c}{\sqrt{1+\left( \frac{\alpha^\prime s}{R^2 G_o^2}\right)^2}}, 
\quad F_{01}^c = \frac{2\pi R^2}{s} \frac{1}{G_o^2 (2\pi \alpha^\prime)^2}
\end{equation}
In the limit $\alpha^\prime\to 0$ the electric field attains its critical
value $F_{01}^c$, but the resulting theory on the 3-brane is not a field 
theory, which would be non-unitary \cite{tom}.
Instead it turns out to be a open string theory with non-commutativity
in the 0-1 plane (NCOS). This can be seen most easily by trading the electric
flux on the 3-brane for a non-zero $B^{NS}_{01}$ in the bulk.

We want to identify the non-interacting states discussed in the previous 
two sections in the NCOS and verify that they do not interact. 
The states are easy to identify. Since they correspond to fundamental strings
with winding and momentum in the compact directions in the string description
of the non-commutative $U(1)$, they should turn into D-strings with
winding and momentum in the S-dual picture.

We have to consider two different wound D-strings, D-strings with $w^3$ units
of winding in the 3-direction and $s w^3$ units of momentum in 2 and states with
winding number $w^2$ in 2 and $-s w^2$ units of momentum in 3. 
For the purpose of counting the unbroken supersymmetries we can introduce
new coordinates
\begin{equation}
\left( \begin{array}{c} x_2^\prime \\ x_3^\prime  \end{array} \right)  =
\frac{1}{(w^2)^2 + (w^3)^2}
\left( \begin{array}{cc} w^3 & -w^2 \\ w^2 & w^3 \end{array} \right) 
\left( \begin{array}{c} x_2 \\ x_3  \end{array} \right).
\end{equation}
An arbitrary superposition of the D-string states is represented as a state
with one unit of winding in the $3^\prime$ direction and $-s$ units of momentum
in the $2^\prime$ direction. In the subsequent analysis we will drop the
primes and only consider states with $(n_2,w^3)=(-s,1)$.

It is very difficult to show that D-string states decouple from the dynamics
of the open strings of the NCOS, but we can provide some circumstantial
evidence that our identification of the decoupled heavy states is correct.
We can compute the masses of these D-string states and we can show that
they are 1/4 BPS states in the presence of the 3-brane with electric flux. 

To address the issue of supersymmetry we can ignore the background RR B-field. 
It shifts the masses of some states, but does not
modify the supercharges preserved by the individual branes. One way to see
this is to note that the RR B-field can be gauged away locally. Unlike in the
case of an NS B-field, this transformation does not induce any fieldstrenghts
on the brane, so the unbroken supersymmetries will not depend on the 
RR B-flux through the torus. 

So, for the purpose of counting the unbroken supersymmetries we set
$B^{RR}_{\mu\nu}=0$. We emphasize that this is a computational tool and that
the physics of the D-string 
does depend on the values of the RR B-field. To construct the
unbroken supersymmetries we need to compute the velocity of a D-string
with momentum $p =-s/R$ in the 2-direction. The canonical momentum can
be computed from the Lagrangian for a D-string written in terms of the open
string quantities 
\begin{equation}\label{action}
L  = \frac{2\pi R}{2\pi\alpha^\prime \hat{G}_o^2}
\sqrt{-\hat{G}_{00}\hat{G}_{33}} \sqrt{1-v_2^2},
\end{equation}
where $v_2$ is the velocity in the 2 direction and
$\hat{G}_{AB} = 2\pi \alpha^\prime/ (G_o^2 \theta) \eta_{AB}$
\footnote{Ref. \cite{sdual} uses the symbol $\tilde{G}_{AB}$ for this metric.}.
Equating the canonical momentum with $p =-s/R$, solving for $v_2$ and 
using $\theta = 2\pi R^2/s$ we find $v_2=1/\sqrt{2}$. 

One way to count the supersymmetries preserved by $s$ fundamental
strings in directions
01 bound to a D3-branes in 0123, and a D-string with one unit of
winding in 3 and $-s$ units of momentum in $2$ is to use T-duality of
type IIB string theory to convert all objects into D-branes. To this end we
compactify all spatial directions of the 3-brane and T-dualize the 1 and 3
directions.  This turns the original D-string into another D-string in 01
with $-s$ units of momentum in 2. The original 3-brane turns into a D-string
in 02 with $s$ units of momentum in 1. We can count the supersymmetries of
this configuration using standard techniques \cite{book}. The only modification
arises from the relative motion of the two branes, which can be taken into 
account by boosting the supercharges appropriately. Using the Born-Infeld
action it is a simple matter to check that the D-string in 02 with $s$ units of
momentum in 1 moves with $v_1=-1/\sqrt{2}$ and we already computed
$v_2=1/\sqrt{2}$ above. The supercharges left unbroken by the two D-strings
separately are
\begin{equation}
\label{cond}
Q_L + \Lambda_\frac{1}{2}(v_1)\Gamma^\perp_2
\Lambda^{-1}_\frac{1}{2}(v_1) Q_R,
\quad
Q_L + \Lambda_\frac{1}{2}(v_2)\Gamma^\perp_1
\Lambda^{-1}_\frac{1}{2}(v_2) Q_R,
\end{equation}
where $\Lambda_\frac{1}{2}(v_i)$ is the matrix for a boost in the $i$
direction with parameter $v_i$ acting on spinor indices, and
$\Gamma^\perp = \prod_{\mu\in\perp} \Gamma\Gamma^\mu$ is obtained by taking the
product over all transverse directions (see \cite{book} for details).

By explicit calculation we find that the configuration of two D-strings
we consider here preserves a quarter of the supersymmetries. Note that this
is true only if the D-strings carry $s$ units of momentum with the appropriate
signs. In all other cases the two conditions in \eq{cond} are incompatible
and the superposition of these D-strings preserves no supersymmetry.
This does not mean that they cannot relax into a supersymmetric ground state,
but it does imply that the system can lower its energy. Such an instability
is usually signaled by a tachyon in the open string spectrum. The main 
observation here is that for states with $s$ units of momentum there is 
no tachyon and the superposition of the two D-strings is a 1/4 BPS state. 
Undoing the two T-dualities we conclude that the D-string states we consider
here form a marginally bound state with the 3-branes with $s$ units of electric
flux, provided they have compact momenta that are multiples of $s$. All other
D-string states have an instability against dissolving in the 3-brane. 

Having counted the supersymmetries preserved by this configuration we are now
in position to compute the mass of these D-string states. The mass does 
depend on the RR flux through the torus.  Consequently we restore the RR B-field
to its original value $B^{RR}_{23}=-1/\theta$ and add the Chern-Simons term
\begin{equation}
L_{cs} = -2\pi R B^{RR}_{23} v_2
\end{equation}
to the Lagrangian of the D-string \eq{action}. Computing the canonical momentum 
from the complete action and solving for the velocity of the D-string now yields
$v_2=0$. Using this result we readily verify from \eq{action} that the mass
of a  D-string with one unit of winding and $s$ units of momentum is $M= s/R$
as expected. We emphasize that these
states have the correct mass only if the RR B-field is included. Setting these
fields to zero is valid in the non-compact case, but if any of the NCOS 
directions are compactified, it is essential to retain both the NS and the
RR B-fields.

While it is straightforward to show that the D-string states discussed above are
1/4 BPS states as expected from the dualities, it is much more involved to 
show that they do not interact. To show that these states decouple from the
NCOS degrees of freedom, we would have to show that the amplitude for two 
NCOS strings to scatter into these D-string states vanishes. This is a 
non-perturbative calculation and as such it is beyond out present capabilities. 
We would also need to verify that  the scattering amplitude for two of
these D-string states in the presence of the 3-brane with $s$ units of
magnetic flux vanishes. Both of these computations are very difficult and we
will not attempt them here. 

\acknowledgements 
It is a pleasure to thank Micha Berkooz, Igor Klebanov, Kostas Skenderis,
Washington Taylor, and especially Ori Ganor for helpful conversations. 
Thanks also to the Aspen Center for Physics where the bulk of this work
was done. 
This work was supported in part by DOE grants \#DF-FC02-94ER40818 and
\#DE-FC-02-91ER40671.

{\tighten 
 
} 
\end{document}